\documentclass[aps,epsfig,float,twocolumn,prl]{revtex4}

\usepackage{graphicx}
\def\<{\langle}
\def\>{\rangle}
\def\i{\mathrm{i}}

\begin{document}
\draft
\title{Local edge modes in doped cuprates with checkerboard polaronic heterogeneity}
\author{Eiji Kaneshita, Ivar Martin, and Alan R. Bishop}
\address{
Los Alamos National Laboratory, Los Alamos, NM 87545}
\date{\today}
\begin{abstract}
We study a periodic polaronic system, which exhibits a nanoscale superlattice structure, as a model for hole-doped cuprates with checkerboard-like heterogeneity, as has been observed recently by scanning tunneling microscopy (STM).
Within this model, the electronic and phononic excitations are investigated by applying an unrestricted Hartree-Fock and a random phase approximation (RPA) to a multiband Peierls-Hubbard Hamiltonian in two dimensions.
\end{abstract}
\maketitle

Recent scanning tunneling microscopy (STM) studies of a Cu oxychloride\cite{STM,STM2}, as well as some other doped cuprates\cite{cb,cb2,cb3,cb4,yazdani} that exhibit high-temperature superconducting behavior, has revealed that doped holes can drive mesoscopic domains of novel nanoscale electronic modulation at low doping ($x \sim \frac{1}{16}$), that is, ``checkerboards", rather than a simple stripe modulation.
The checkerboard domains, according to analysis of the experimental data\cite{STM,STM2}, consist of $(4\times4)$-unit-cell plaquettes; the plaquette size is essentially insensitive to the doping level, presumably locking to this commensurate periodicity with discommensuration defects.

However, it is not clear where doped holes are located.
If the holes lie along the grid of a checkerboard,
the linear hole concentration on the grid would be extremely dilute for underdoped cases.
It is unlikely, taking into account that the system is an insulator at such a low doping level, that holes compose a 4-site-periodic checkerboard and are diluted throughout the whole plane like a metal.
Thus it seems more likely that each doped hole is localized, and forms a ``polaron" with associated lattice and spin distortions.

Here, therefore, we study a periodic polaronic structure as a model for hole-doped cuprates with checkerboard-like superlattice heterogeneity.
Our aim is to determine the spectroscopic signatures of such checkerboards, and contrast them with signatures of stripe segments.
We expect that the results of this work will aid in the analysis and interpretation of recent experimental data of angle-resolved photoemission spectroscopy (ARPES), nuclear quadrupole resonance (NQR), electron paramagnetic resonance (EPR), and the dynamical structure factor $S(q,\omega)$.\cite{shen,SQW,SQW2,NQR,NQR2,NQR3,EPR}

There have been several attempts to describe cuprate superconductors as a system with stripe modulation.
We\cite{martin} previously investigated the phonon mode structure in the presence of stripes via Hartree-Fock and RPA analysis in a three-band Peierls-Hubbard model, and reported that, owing to the stripe inhomogeneity, local edge (interface) modes of oxygen lattice vibrations are generated with characteristic energies around 15-45 and 70 meV; these energies are similar to those where the electronic band dispersion has a ``kink" in ARPES\cite{shen} and where anomalous inelastic neutron scattering has been identified\cite{mcqueeney,mcqueeney2}.
These modes are also accompanied by correlated electronic excitations in both magnetic and charge channels\cite{martin}.
The dispersive phonon branch at around 15-40 meV is due to oxygen vibrations \textit{along} a vertical stripe and is only present for the vertical/horizontal stripe orientation.
The 70 meV mode is from transverse oxygen vibrations and is present for all stripe orientations.

In this study, we adopt a system with periodic polarons forming a $4\times4$ superlattice for a $\frac{1}{16}$-hole-doping case, and investigate the local electronic and phonon modes to complement and discriminate from the stripe cases.
Although the hole-doping level here is extremely low ($x=\frac{1}{16}$), we compare the results with those of the stripe cases with a different doping level ($x=\frac{1}{3}$), because it is numerically difficult to calculate such a large system.
However, this comparison is justified, because the energy of the phonon edge modes in the stripe cases, being a local property, is essentially insensitive to the doping level\cite{martin}.

To model the CuO$_2$ planes of doped cuprates, we use a two-dimensional (2D) three-band extended Peierls-Hubbard Hamiltonian, which includes both electron-electron and electron-phonon interactions\cite{yone,yi}:
\begin{eqnarray}\label{eq:H0}
H_0 = \sum_{<ij> \sigma} t_{pd}(u_{ij})
(c^\dagger_{i\sigma}c_{j\sigma}+ H.c.)
+\sum_{i,\sigma} \epsilon_i(u_{ij})
c^\dagger_{i\sigma}c_{i\sigma} \nonumber\\
+\sum_{<ij>} \frac{1}{2} K_{ij}u^2_{ij}
+\tilde{\sum_{i, j,\sigma, \sigma'}}\frac{U_{ij}}{2}n_{i\sigma}n_{j\sigma'}.
\end{eqnarray}
Here, $c^\dagger_{i\sigma}$ creates a hole with spin $\sigma$ on site $i$;
each site has one orbital (d$_{x^2-y^2}$ on Cu, or O p$_x$ or p$_y$ on O).
The Cu (O) site electronic energy is $\epsilon_d$ ($\epsilon_p$).
$U_{ij}$ represents the on-site Cu (O) Coulomb repulsion, $U_d$ ($U_p$), or the intersite one, $U_{pd}$, and the summation $\tilde{\sum}$ is taken, except for the case of $(i,\sigma)=(j,\sigma')$.
The electron-lattice interaction causes modification of the Cu-O hopping strength through oxygen displacement $u_{ij}$:
$t_{pd}(u_{ij}) = t_{pd} \pm \alpha u_{ij}$, where $+(-)$ applies if the Cu-O bond shrinks (stretches) for a positive $u_{ij}$;
it also affects the Cu on-site energies $\epsilon_d(u_{ij}) = \epsilon_d + \beta\sum_j{(\pm u_{ij})}$,
where the sum runs over the four neighboring O ions.
Other oxygen modes (buckling, bending, etc) are assumed to couple to electron charge more weakly and are neglected for simplicity, but can be included as necessary within the same approach.
We use the following set of model parameters\cite{martin,yone}:
$\epsilon_p-\epsilon_d = 4.4$ eV, $U_d =11$ eV, $U_p = 3.3$ eV, $U_{pd} = 1.1$ eV, and $K = 38.7$ eV/{\AA$^2$}, $\alpha = 5.2$ eV/\AA, $\beta = 1.2$ eV/\AA, with $t_{pd} = 1.1$ eV.
To approximately solve the model, we use the unrestricted Hartree-Fock approximation combined with an inhomogeneous generalized RPA for linear fluctuations of lattice, spin or charge\cite{yone} in a supercell of size $N_x\times N_y$ with periodic boundary conditions.

The output of the calculation here is the inhomogeneous Hartree-Fock ground state and the linearized fluctuation eigenfrequencies and eigenvectors.
From the \textit{phonon} eigenmodes, we calculate the corresponding neutron scattering cross section:
\begin{eqnarray}
S(\mathbf{k},\omega) =\int dt\, e^{-i\omega t}
\sum_{l l^\prime}{\langle e^{-i \mathbf{k R}_l(0)}
e^{i\mathbf{k R}_{l'}(t)}\rangle},
\end{eqnarray}
where $\mathbf{R}_{l}(t) =  \mathbf{R}_{l}^0 + \mathbf{d}_l + \mathbf{u}_l(t)$ is the position of the $l$-th oxygen atom expressed in terms of the location of the unit cell origin $\mathbf{R}_{l}^0$, the position within the unit cell $\mathbf{d}_l$, and the time-dependent vibrational component $\mathbf{u}_l(t)$.
For phonon modes with $\mathbf{u}_l(t)$ oriented along the corresponding metal-oxygen bonds, on the O$_x$ sublattice $\mathbf{d}_l = \frac a2 \hat{x}$ and $\mathbf{u}_l\equiv x_l \hat{x}$, and on the O$_y$ sublattice $\mathbf{d}_l = \frac a2 \hat{y}$ and $\mathbf{u}_l\equiv y_l \hat{y}$.
The scalar displacements can now be expressed in terms of the normal modes $z_n$ as $x_l(t) = \sum_n{\alpha_{x_l,n}z_n(t)}$ and $y_l(t) = \sum_n{\alpha_{y_l,n}z_n(t)}$.
performing first- order expansion in the oxygen displacements, we obtain
\begin{eqnarray}
S(\mathbf{k},\omega) &=& \sum_n
\Big[k_x^2|\alpha_{\mathbf{k},n}^x|^2
+ k_y^2|\alpha_{\mathbf{k},n}^y|^2\\\nonumber
&& + k_x k_y(e^{-\i(k_x-k_y)a/2}
\alpha_{\mathbf{k},n}^x
\alpha_{-\mathbf{k},n}^y + c.c.)\Big]\\\nonumber
&&\times\frac{\hbar}{2m\omega_n}
[(1+n_B)\delta(\omega-\omega_n) + n_B\delta(\omega+\omega_n)].
\end{eqnarray}
Here, $\alpha_{\mathbf{k},n}^x  = \sum_l{e^{-\i\mathbf{k R}_l^0}\alpha_{x_l,n}}$, and $n_B = (e^{\omega_n/T}-1)^{-1}$ is the thermal population of the phonon mode $n$. This is a generalization of the usual neutron scattering intensity expression\cite{lovesay} for the case of phonons with a larger real space unit cell.
We plot $S(\mathbf{k},\omega)/|\mathbf{k}|^2$ for \textbf{k}-directions sampling longitudinal modes, consistent with the common experimental convention.

To investigate the neutral \textit{electronic} excitations, we calculate the spectral function:
\begin{eqnarray}
\sum_n |\, \langle \Psi_0| \mathcal{O}(\mathbf{k})|\Psi_n \rangle \,|^2\,\delta(\omega-E_n-E_0),
\end{eqnarray}
where $|\Psi_0\rangle$ ($|\Psi_n\rangle$) is the Hartree-Fock ground (excited) state whose energy is represented by $E_0$ ($E_n$), and $\mathcal{O}(\mathbf{k})$ is an operator, e.g. spin $\mathbf{S}(\mathbf{k})$ or charge $n(\mathbf{k})$, summed over Cu- and O-sites:
\begin{eqnarray}
\mathcal{O}(\mathbf{k})=\hspace{-0.3cm}\sum_{
\scriptstyle \lambda\,=\,\mathrm{Cu},
\atop \scriptstyle\,\mathrm{O}_x,\,\mathrm{O}_y}
\hspace{-0.2cm}
\mathcal{O}^{(\lambda)}(\mathbf{k})\,e^{-\i \mathbf{kd}_\lambda},
\hspace{0.3cm}
\mathbf{d}_\lambda=\left\{
\begin{array}{lr}
\hat{0} &\mbox{for $\lambda=\mathrm{Cu}$}\\
\frac{a}{2} \hat{x}&\mbox{for $\lambda=\mathrm{O}_x$}\\
\frac{a}{2} \hat{y}&\mbox{for $\lambda=\mathrm{O}_y$}
\end{array}
\right..
\end{eqnarray}
The effect of an infinitesimal external field corresponding to the excitation $\Psi_n$ can be represented by the change of an observable $\langle\mathcal{O}\rangle$ in the state $\Psi=\Psi_0 + \eta \Psi_n$ ($|\eta|\ll1$):
\begin{eqnarray}
\langle\mathcal{O}\rangle\simeq\langle\mathcal{O}\rangle_0+\delta\langle\mathcal{O}\rangle_n,\\
\delta\langle\mathcal{O}\rangle_n\propto\langle\Psi_0| \mathcal{O} |\Psi_n\rangle,
\end{eqnarray}
where $\langle\mathcal{O}\rangle_0$ is the expectation value with respect to the ground state.

\begin{figure}[htbp]
\begin{center}
\includegraphics[width = 0.9\linewidth]{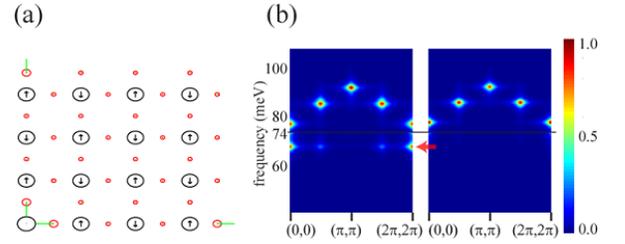}
\caption{(a) The ground-state configuration given by mean-field Hartree-Fock calculation.
Black and red circles represent Cu and O sites, respectively.
The radius of the circle is proportional to the corresponding site hole density.
Arrows centered on circles show the magnitude and direction of spin.  The green lines originating from O sites indicate the magnitude and direction of equilibrium O displacements in the presence of a polaron.
(b) The phonon spectra calculated by means of RPA for the doped polaronic case (left) and the undoped AF case (right); here the intensity is normalized by the maximum value, and magnified 4-fold below 74 meV.
In the polaronic case (left), there is an extra branch (indicated by the red arrow), which is from a local vibration mode around a polaron, near 68 meV.
} \label{fig:ground}
\end{center}
\end{figure}

We first show the ground state configuration given by the Hartree-Fock calculation (Fig.~\ref{fig:ground}(a)).
The ground state has one polaron per unit cell of the polaronic $4\times4$ superlattice.
In the case of two holes per unit supercell, the extra hole is preferentially located at the central site of the one-hole-polaron supercell.
Because of the symmetry, the central sites are likely to be important in understanding both the phonon mode and electronic excitation even in the one-hole case.

The phonon spectra are then obtained by real-space RPA (Fig.~\ref{fig:ground}(b)).
In these spectra, a dispersive branch occurs at 77-92 meV in both the doped and undoped cases.
This branch is essentially insensitive to doping.
However, doping causes an extra dispersionless branch near 68 meV.
This branch corresponds to a local mode --- the ``shape mode" --- due to the heterogeneity of the system with sharp interfaces, i.e., the local vibrations of the oxygens around a polaron.
These results are very similar to those found in the stripe case\cite{martin}.
However, note again that, in the vertical stripe case, there are two branches of local phonons at the stripe that correspond to oxygen oscillations with directions parallel and perpendicular to the stripe, while there is only one local phonon branch in the polaronic or diagonal stripe cases.

\begin{figure}[htbp]
\begin{center}
\includegraphics[width = 0.9\linewidth]{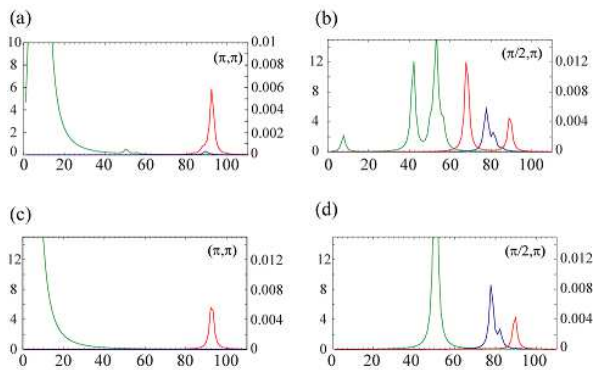}
\caption{The excitation spectra for charge (red), longitudinal spin (blue) and transverse spin (green) channels given by real-space RPA. The spectra for charge and longitudinal spin are scaled on the right axis, and that for transverse spin is scaled on the left axis.
(a) and (b) are for the polaronic case; (c) and (d) for the undoped case.
} \label{fig:spectra-el}
\end{center}
\end{figure}

\begin{figure}[htbp]
\begin{center}
\includegraphics[width = 0.9\linewidth]{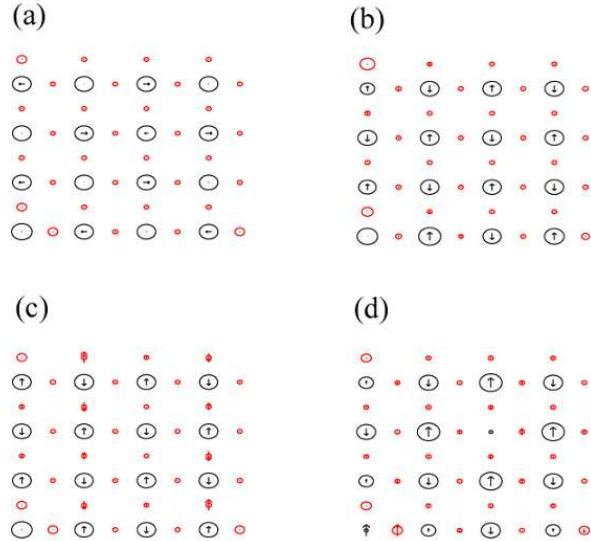}
\caption{The excited-state spin eigenvectors for (a) $E\sim51$ meV, (b) $E\sim68$ meV,  (c) $E\sim77$ meV and (d) $E\sim89$ meV.
The arrows in (a) represent $S_x$, while those in (b)-(d) represent $S_z$.
} \label{fig:excitation}
\end{center}
\end{figure}

Next, we calculate the electronic excitations within the same real-space RPA.
The transverse \textit{spin} mode has a non-zero-gap excitation ($\sim$ 8 meV), corresponding to AF spin waves, at several k-points: ($\pi,\pi$), ($\frac{\pi}{2},\pi$), ($0,\pi$), etc. This is because the polaronic configuration cannot be expressed by a single \textbf{k}.
The excitations at ($\pi,\pi$) in this channel has not only the $\sim$ 8 meV peak but also peaks at $\sim$ 51 meV and $\sim$ 57 meV (Fig.~\ref{fig:spectra-el} (a)) for sub-spin waves (i.e., higher energy transverse spin excitations generated from low-energy spin waves, as shown in Fig.~\ref{fig:excitation} (a)). These peaks also exist at other \textbf{k}-points.
The excitation modes corresponding to the peak at $\sim$ 51 meV are again similar to those found for vertical stripes; however, the homogeneous system does not have this mode at $(\pi,\pi)$.
This mode occurs by the folding back of the spin wave branch.
It depends on the symmetry of the superlattice; e.g. the two-hole system does not exhibit the sub-spin waves at $(\pi,\pi)$.

Furthermore, the same kind of excitation appears in the undoped case (Fig.~\ref{fig:spectra-el} (d)), at other than $(\pi,\pi)$.
However, this excitation has a single peak (unlike the polaron case) whose energy depends on the momentum.
This is not surprising because this is the branch of the spin wave excitation originating at ($\pi,\pi$).

Finally, the \textit{charge} mode at ($\pi,\pi$) has a well-defined peak near 92 meV (Fig.~\ref{fig:spectra-el} (a)), coupled with the corresponding phonon excitation.
This mode shows the typical $(\pi,\pi)$ charge oscillation at the Cu sites, as in the case of undoped AF.
The longitudinal spin mode at ($\frac{\pi}{2},\pi$) has well-defined peaks at $\sim$ 77 meV and $\sim$ 81 meV (Fig.~\ref{fig:spectra-el} (b)).
These modes show the oscillation of $S_z$ at the oxygen sites, as in the case of undoped AF.
However, unlike the undoped case, a weak peak also exists at $\sim$ 89 meV at ($\pi,\pi$) in the charge and the longitudinal spin channels.
One of the excited states corresponding to this energy at $\sim$ 89 meV is shown in Fig.~\ref{fig:excitation} (d).

At the characteristic energy of 68 meV, where the local phonon mode exists, the charge channel exhibits a major difference between the polaronic and the undoped AF cases (Fig.~\ref{fig:spectra-el} (b)).
The charge mode of this energy yields well-defined peaks in the excitation spectra at $(0,\pi)$, $(0,\frac{\pi}{2})$, $(\frac{\pi}{2},\pi)$ and the symmetric equivalents of those points, and the intensity is the highest at $(\frac{\pi}{2},\pi)$ ---and also at the symmetrically equivalent points--- in \textbf{k}-space.
From the analysis of the excited state corresponding to this mode, we find that the charge on the polaron undergoes local oscillations of its amplitude, coupled with the local phonon mode (the so-called half-breathing mode \cite{mcqueeney,mcqueeney2}).

In summary, we have studied a periodic $4\times4$-superlattice polaronic ground state as a model of the nanoscale heterogeneous structure recently observed in some hole-doped oxychlorides and other hole-doped cuprates.
We investigated the characteristic electronic and lattice excitations relative to that ground state.
The inhomogeneity of the system causes excitations distinctly different from those in the homogeneous case.
The major new excitations occur in the charge channel ($\sim$ 70 meV mode), and are coupled with local half-breathing phonon modes, as well as in all the stripe cases.
However, the oscillation patterns of this excitation are, compared with those directed along the stripe in the vertical stripe case, localized at the polaron site.
Assuming the system possesses a vertical stripe (or stripe segment) structure, we expect to find two kinds of excitations at around 15-45 meV and 70 meV.
On the other hand, if the polaronic superlattice structure exists, then there are only excitations at around 70 meV.
A diagonal stripe case is similar to the polaronic case for charge excitations; however, in the transverse spin channel, there is an additional excitation at $\sim$ 50 meV in the $4\times4$ polaronic case.
The recent analysis of ARPES data supports the presence of these low-frequency local (edge) mode excitations, which are characteristics of polaronic inhomogeneity existing as either checkerboard or stripe (segment) domains.
As we predict, these assignments are correlated with inelastic neutron scattering (for spin and charge) \cite{mcqueeney,mcqueeney2} showing the existence of the modes at around 40 and 70 meV.
NQR\cite{NQR,NQR2,NQR3} and EPR\cite{EPR} data are likewise consistent with the assignment.
The correlation among spin, charge and lattice signatures of these local (interface) modes will be essential for their definitive assignments and awaits further experimental study.

 Finally, we emphasize that similar signatures of nanoscale
electron/lattice inhomogeneity can be expected below the polaron formation
temperature in many doped transition metal oxides
(bismuthates, manganites, nickelates, etc.) and related strongly correlated
electronic materials. It will be important to compare these cases and
identify which signatures correlate with the appearance of
superconductivity.

We thank Dr. Zhi Gang Yu for valuable consultations on the Hartree-Fock and RPA techniques.
This work was supported by the U.S. DOE.


\begin{thebibliography}{99}
\bibitem{martin} I. Martin, E. Kaneshita, R. J. McQueeney, Z. G. Yu and A. R. Bishop: to be published in Phys. Rev. B; Note that the calculation in this article is not for $x=\frac{1}{16}$ but for $x=\frac{1}{3}$.
\end{thebibliography}
\end{document}